\documentclass[superscriptaddress, aps, prx, reprint, showkeys, floatfix]{revtex4-2}

\usepackage{graphicx}
\usepackage{bm}

\usepackage{siunitx}
\usepackage{chemformula}
\usepackage{enumitem}
\usepackage{xcolor}
\usepackage{lipsum}

\newcommand{\DCN}{Dresden Center for Nanoanalysis, cfaed, TUD University of Technology Dresden, 01069 Dresden, Germany.}
\newcommand{\IFW}{Leibniz Institute for Solid State and Materials Research Dresden, 01069 Dresden, Germany.}
\newcommand{\IFMP}{Institute for Solid State and Materials Physics, TUD University of Technology Dresden, 01069 Dresden, Germany.}
\newcommand{\IoP}{Institute of Physics, Academy of Sciences of the Czech Republic, Cukrovarnick\'a 10, 162 00 Praha 6, Czech Republic}
\newcommand{\MPI}{Max Planck Institute for Chemical Physics of Solids, 01187 Dresden, Germany.}

\begin{document}


\title{In-situ monitoring the magnetotransport signature of topological transitions in the chiral magnet \ch{Mn_{1.4}PtSn}} 

\author{Andy Thomas}
 \email{a.thomas@ifw-dresden.de}
 \affiliation{\IFMP}
 \affiliation{\IFW}
\author{Darius Pohl}
\email{darius.pohl@tu-dresden.de}
 \affiliation{\DCN}
\author{Alexander Tahn}
 \affiliation{\DCN}
\author{Heike Schl\"orb}
 \affiliation{\IFW}
 \author{Sebastian Schneider}
 \affiliation{\DCN}
\author{Dominik Kriegner}
 \affiliation{\IoP}
 \affiliation{\IFMP}
\author{Sebastian Beckert}
 \affiliation{\IFMP}
\author{Praveen Vir}
 \affiliation{\MPI}
\author{Moritz Winter}
 \affiliation{\MPI}
 \affiliation{\DCN}
\author{Claudia Felser}
 \affiliation{\MPI}
\author{Bernd Rellinghaus}
 \email{bernd.rellinghaus@tu-dresden.de}
 \affiliation{\DCN}

\date{\today}

\begin{abstract}
Emerging magnetic fields related to the presence of topologically protected spin textures such as skyrmions are expected to give rise to additional, topology-related contributions to the Hall effect. In order to doubtlessly identify this so-called topological Hall effect, it is crucial to disentangle such contributions from the anomalous Hall effect. This necessitates a direct correlation of the transversal Hall voltage with the underlying magnetic textures. We utilize a novel measurement platform that allows to acquire high-resolution Lorentz transmission electron microscopy images of magnetic textures as a function of an external magnetic field and to \emph{concurrently} measure the (anomalous) Hall voltage in-situ in the microscope on one and the same specimen. We use this approach to investigate the transport signatures of the chiral soliton lattice and antiskyrmions in \ch{Mn_{1.4}PtSn}. Notably, the observed textures allow to fully understand the measured Hall voltage without the need of any additional contributions due to a topological Hall effect, and the field-controlled formation and annihilation of anstiskyrmions are found to have \emph{no} effect on the measurend Hall voltage.
\end{abstract}

\keywords{In-situ TEM, Hall effect, Lorentz TEM, Micromagnetic simulations}

\maketitle

%
%
The Hall effect in solids, discovered by Edwin H. Hall in 1879 \cite{Hall:1879}, describes the formation of a voltage perpendicular to the applied current and magnetic field. In materials with broken time-reversal symmetry (e.g. ferromagnets), the Hall resistivity consist of the ordinary Hall effect (OHE) caused by the Lorentz force as well as the anomalous Hall effect (AHE) resulting from different scattering mechanisms like intrinsic, skew, and side jump scattering \cite{Nagosa:2010}. Therefore, the measured Hall resistivity is often empirically described by
\begin{equation}
    \rho_{xy} = \rho_{xy}^o \mu_0 H + \rho_{xy}^a \mu_0 M(H) + \rho_{xy}^t
    \label{eq:hall}
\end{equation}
with the permeability of free space $\mu_0$, the external magnetic field $H(H)$ and the magnetization of the sample $M(H)$ \cite{Porter:2014}. The first term describes the OHE and the second one accounts for the AHE. Even though the origin of the AHE is still subject of an ongoing debate, further contributions to the Hall resistivity, which are not proportional to the magnetization, have become of increasing research interest. Consequently, the third term designates additional contributions from the Berry phase leading to a topological Hall effect (THE). For example, the non-vanishing Berry phase could be caused by topologically protected magnetic textures such as skyrmions and its relatives \cite{Huang:2012,Bruno:2004,Neubauer:2009,Kanazawa_2011,Huang_2012,Shiomi_2013,Liu_2017,Liu_2018,Li_2013,Lee_2009,Li_2017,Rana_2016,Liang_2015,Matsuno_2016,Ohuchi_2015,Gallagher_2017,Li_2018,Swekis_2019}.

If the THE shall be extracted from Hall measurements, the OHE and AHE have to be subtracted from the Hall signal $\rho_{xy}$. The OHE is usually small and a linear subtraction is rather straightforward, but the subtraction of the large AHE requires precise knowledge of $M(H)$ in the same specimen. If this data is not available, the interpretation of THE-like features in magnetotransport experiments on materials with topological structures such as (anti)-skyrmions relies on magnetometry data, micromagnetic simulations  and/or magnetic imaging techniques \cite{Li_2013,Fert:2017,Zhang:2020}. This has several inherent disadvantages: Micromagnetic simulations require precise knowledge of numerous material parameters, at least part of which are unknown or merely estimated. In addition, magnetometry data and/or magnetic images are usually not obtained from identical specimen, or the resolution in the magnetic images does not allow to identify the topology of the underlying textures \cite{Winter2022}. 

In a more sophisticated approach to extract the THE, the controversial procedure of subtracting the AHE via $M(H)$ can be avoided by a simultaneous measurement of the Hall and Nernst effects \cite{schlitz2019}. This, however, requires to also measure the (topological) Nernst effect thereby necessitating an increased experimental effort, which was not yet demonstrated for (anti)-skyrmions. 

Skyrmions and even more complex magnetic textures are frequently studied using advanced magnetic imaging techniques such as Lorentz transmission electron microscopy (LTEM) or electron holography. However, these techniques are used to identify topological texture like (anti)skyrmions in different materials as a function of magnetic field and temperature \cite{schneider:2018,Stolt:2019,Zhang:2020,Peng:2020,wolf:2022}. Besides determining the topological charge of the magnetic texture, also dynamic studies can be performed \cite{Legrand:2017,Hrabec:2017,Zhang:2020}. In transport experiments, both, magnetoresistance \cite{Tang:2023} and Hall measurements \cite{Neubauer:2009} can be performed to electrically detect skyrmions and the THE is considered to be a strong evidence for the existence of skyrmions in the material \cite{Huang:2012}. However, Gerber considered two independent contributions to the AHE in equation \ref{eq:hall} and a subtraction of only one may be erroneously misinterpreted as a THE \cite{Gerber:2018}. 

\begin{figure}[tbp]
\includegraphics[width=7.0cm]{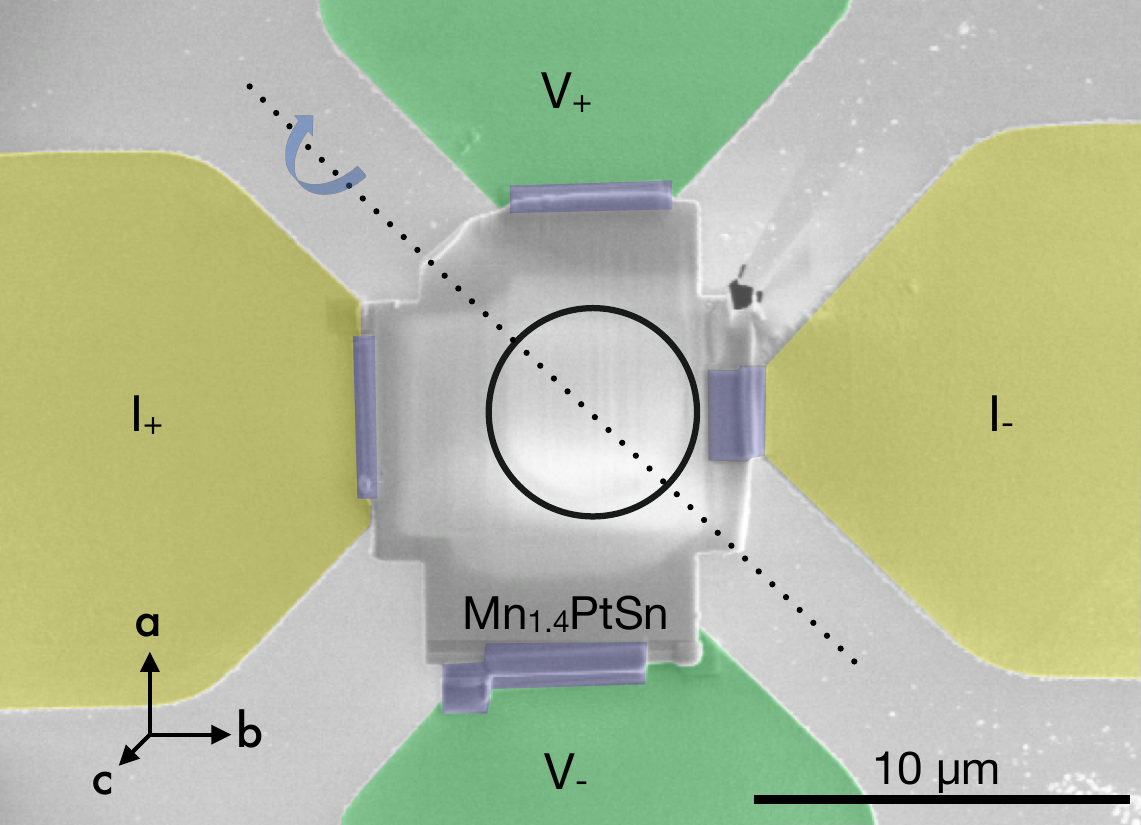}
\caption{Plan view SEM image of the contacted \ch{Mn_{1.4}PtSn} lamella on a commercial \ch{Si} chip with a \ch{Si3N4} window. The lamella is welded to the contact pads of the holder by local ion-beam assisted deposition of tungsten (highlighted in purple) establishing Hall contacts for current supply I (yellow) and voltage measurement V (green) and labelled according to their polarity. A circular hole (indicated by the black circle) was cut in the free-standing SiN membrane at the center of the chip to improve the Lorentz TEM contrast. The dashed line indicates the rotation axis of the holder in the TEM.}
\label{fig:sample}
\end{figure}

Furthermore, in magnetic systems with pronounced sensitivity to dipole-dipole interactions, the stability of magnetic textures depends strongly on the sample geometry. This makes the correlation of magnetotransport and transmission electron microscopy (TEM) data challenging, if not conducted on the identical specimen. Fortunately, modern in-situ TEM holders allow for the feed-through of electrical contacts, which open up the possibility for electrical characterization in-situ in the microscope \cite{Almeida:2020,Romero:2019,Wittig:2017,Tang:2023}. 

Here, we utilize our capabilities of in-situ magnetotransport measurements in a transmission electron microscope \cite{pohl2023} to collect a set of Lorentz-TEM images and corresponding Hall measurements. Using these data, we are able to look into the magnetotransport signature of antiskyrmions in \ch{Mn_{1.4}PtSn}. 

\ch{Mn_{1.4}PtSn} was chosen because it is reported to host antiskyrmions at room temperature. Recent magnetotransport measurements on this material combined with magnetic imaging using the magneto-optical Kerr effect (MOKE) by M.\ Winter et al.\ \cite{Winter2022} had implied the occurrence of additional contributions to the Hall effect at fields just below saturation. Here, sub-micron sized magnetic objects were observed in the MOKE images. Although the field-induced creation of antiskyrmions was predicted in theoretical calculations, neither low-temperature magnetic force microscopy nor the room temperature MOKE images allowed to identify the topology of these circular magnetic nano-objects. Consequently, the origin of this additional contribution to the Hall effect and in particular the nature of the underlying magnetic texture, remained unclear. Here, LTEM provides for the necessary high resolution to clearly identify these nano-magnetic textures.

\begin{figure}[tbp]
\includegraphics[width=\linewidth]{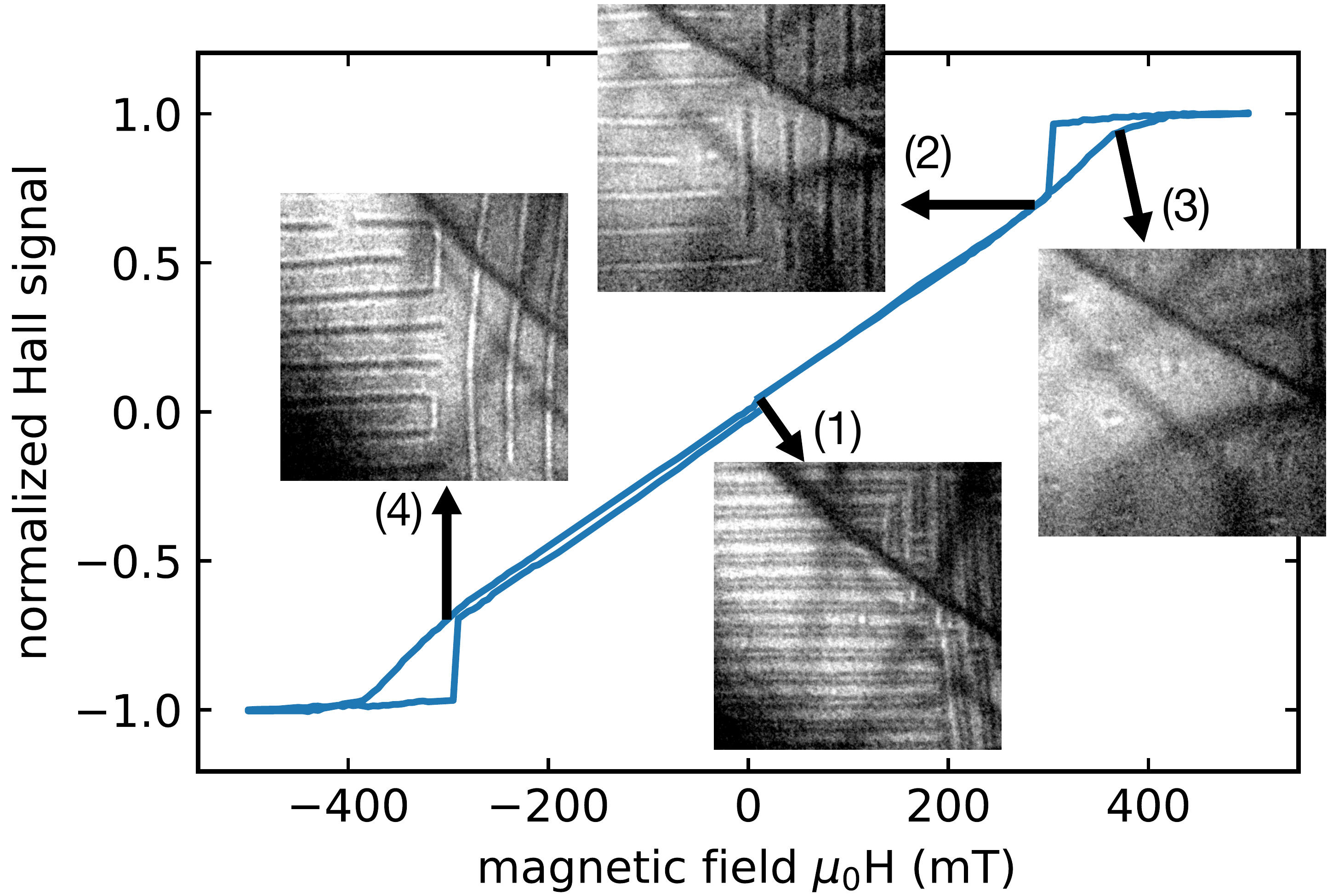}
\caption{Normalized Hall signal of a thin lamella of \ch{Mn_{1.4}PtSn} 
measured in-situ in a transmission electron microscope as a
function of the magnetic field provided by the objective lens.
Concurrently acquired LTEM images (\SI{1.6}{\micro m} squares) depict the respective
magnetic states of the sample and the arrows indicate the corresponding position on the Hall curve. }
\label{fig:Hall_survey}
\end{figure}

\ch{Mn_{1.4}PtSn} is a non-centro\-symmetric, tetragonal compound (space group $I\bar{4}2d$) with D$_{2d}$ symmetry. The lattice parameters $a$ and $c$ are \SI{0.662}{nm} and \SI{1.224}{nm}, respectively. Single crystals of \ch{Mn_{1.4}PtSn} were grown by the flux-growth method using \ch{Sn} flux. Please refer to Vir and coworkers for a detailed description of the single crystals growth and characterizations \cite{Vir2019,vir_anisotropic_2019}. 

From one of these single crystals a TEM lamella was cut using a {\em FEI Helios 660} dual beam focused ion beam (FIB) system. Special care was taken to ensure a homogenous thickness of \SI{100}{nm} across the lamella of approximately 8x14 $\rm \mu m^2$ in size. Commercially available SiN-covered Si chips were used for the in-situ electrical characterization in a {\em Protochips Fusion Select} holder. In order to reduce artifacts in the LTEM images, a circular hole was FIB-cut in a freestanding SiN membrane (without any Si sup
port underneath) at the center of the chip, over which the lamella was placed and electrically connected to the contact pads of the chip using tungsten deposition (cf.\ Fig.\ \ref{fig:sample} for details). 

\begin{figure*}[t]
\includegraphics[width=0.8\textwidth]{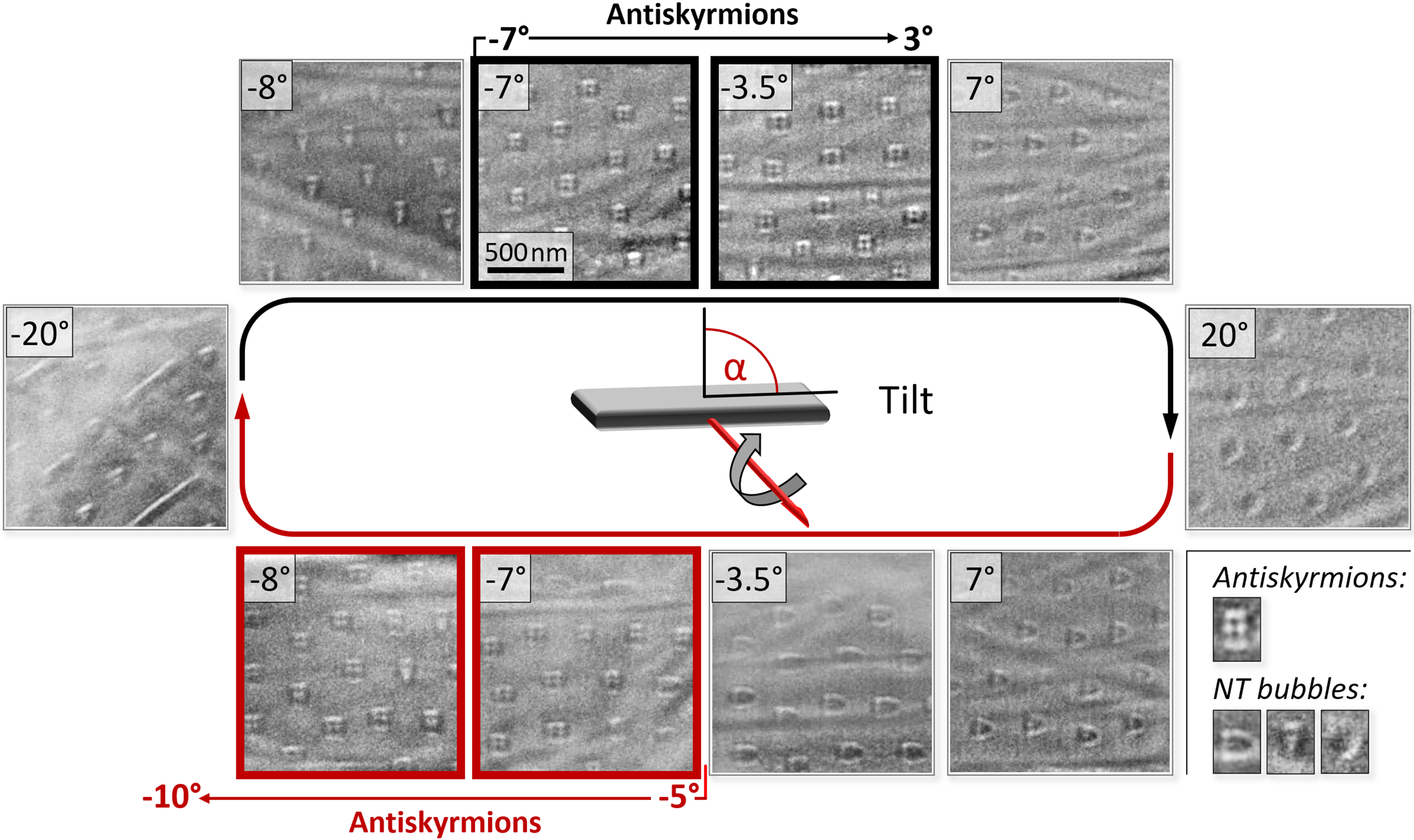}
\caption{Variation of the LTEM contrast upon tilting the sample back and forth from $\rm \ang{-7}$ to $\rm \ang{20}$, $\rm \ang{-20}$, and back to $\rm \ang{-7}$. Antiskyrmions are observed to occur in the ranges $\rm \ang{-7} \leq \alpha \leq \ang{-3}$ and $\rm \ang{-5} \leq \alpha \leq \ang{-10}$ upon tilting forward (black) and backward (red), respectively, non-topological (NT) bubbles occur otherwise. The differences in TEM contrast between antiskyrmions and non-topological bubbles are indicated exemplary in the lower right.}
\label{fig:LTEM-tilt}
\end{figure*}

The in-situ Hall measurements and LTEM investigations were conducted in a {\em JEOL JEM-F200} transmission electron microscope equipped with a cold field emission gun and a {\em GATAN OneView} CMOS camera for fast imaging. The microscope was operated at an acceleration voltage of \SI{200}{kV}. The current through the Hall bar was supplied by a {\em Keithley 2450} source meter, and the transverse Hall voltage was measured by a {\em Keithley 2182A} nanovoltmeter. The excitation of the objective lens to adjust the magnetic field exerted on the sample, the {\em GATAN OneView} camera used for the acquisition of Lorentz-TEM (LTEM) images, the projective lens system of the microscope, and the electrical transport measurements are controlled by Python scripting \cite{PyJEM_GIT,PyJEM_JEOL,Gatan_GMS}. For a detailed description of the in-situ Hall measurement setup please refer to our previous work \cite{pohl2023}.

Figure \ref{fig:Hall_survey} shows the normalized Hall signal of the \ch{Mn_{1.4}PtSn} lamella as a function of the applied magnetic field while a current of \SI{25}{\micro A} is supplied. The selected, concurrently acquired LTEM images indicate that the observed variation of the Hall voltage is accompanied by three distinct magnetic configurations of the sample: (1) a stripe domain pattern with symmetric bright and dark contrasts in different orientational variants at low fields, (2) an increasingly asymmetric stripe pattern with field-driven enhancement of the periodic length at intermediate fields, (3) a mixture of antiskyrmions (AS) and non-topological bubbles (NT) and (4) the same pattern as (2) at negative fields which leads to the expected and observed inversion of the stripe contrast. Upon increasing the field, the stripe features persists up to almost saturation and vanish only at fields, where antiskyrmions and non-topological bubbles, recognized from their typical LTEM contrast \cite{Peng:2020}, occur. At roughly $\mu_0H = \SI{430}{mT}$, any magnetic contrast vanishes indicating that the out-of-plane magnetic saturation of the sample is reached. Upon successively decreasing the magnetic field again, this saturated state is found to persist down to a field of $\mu_0H = \SI{300}{mT}$, where the material re-enters the previously observed asymmetric stripe phase. This re-entrance field coincides precisely with the occurrence of a downward drop of the Hall voltage thereby correlating the hysteretic variation of the magnetic states observed in the LTEM images with the measured hysteresis in the Hall voltage between \SI{300}{mT} and \SI{400}{mT}. (The complete LTEM image series can be found in the extended data set \cite{opara}.)

The increase of the periodic length in the stripe domain patterns is already known from chiral soliton lattices (CSL) \cite{Dzaloshinskii_1964, Izyumov_1984, Kishine_2005}. Togawa and coworkers developed a model to describe the domain pattern of such a CSL as observed with LTEM in mono-axial helimagnets \cite{togawa_chiral_2012}. In the present work, we rather focus on the occurrence and annihilation of antiskyrmions and study their impact on the Hall effect, i.e.\ the field range at approximately \SI{375}{mT} shown in Fig. \ref{fig:Hall_survey} configuration (3).

However, the so far considered out-of-plane magnetic field sweeps are not ideal to investigate the topological Hall effect. The magnetic textures observed in subsequent sweeps at fields just before reaching saturation vary from measurement to measurement: We always observe a mixture of antiskyrmions and non-topological bubbles and their relative amount and positions vary, thereby inhibiting the disentanglement of the topological and anomalous Hall effect. 

L.~Peng and co-workers have shown that by varying the in-plane and out-of-plane components of the applied field through tilting the sample allows to control the relative amount antiskyrmions and bubbles, respectively \cite{Peng:2020}. In order to systematically measure the topological Hall effect of an antiskyrmion lattice, we have therefore performed a series of tilt experiments under a constant magnetic field provided by the objective lens of the microscope. In these experiments, we were able to transform non-topological bubbles into antiskyrmions and vice-versa in a controlled fashion simply by adjusting the tilt angle of the sample and thereby the in-plane field component without changing the lateral position of the object. At every tilt angle, the Hall voltage was measured and an LTEM image was acquired. Post-acquisition alignment of the LTEM images by cross-correlation was used to compensate for any sample drifts during tilting.

In Fig.~\ref{fig:LTEM-tilt} we display representative examples of LTEM images acquired during this tilt series (see Fig.~\ref{fig:sample} for the direction of the tilt axis and the extended data set \cite{opara} for all images).  The sample was initially tilted to \SI{-7}{\degree} in a magnetic field of $\mu_0H_{\perp} = \SI{374}{mT}$ and an electrical current of \SI{50}{\micro A}, to generate an antiskyrmion lattice. The field was chosen, since antiskyrmions were observed to occur in this range in the field sweep experiments. Upon tilting to \SI{+20}{\degree} in steps of \SI{0.5}{\degree}, a non-topological (NT) bubble lattice emerged due to the increased in-plane field component. The sample is then step-wise tilted back to \SI{-20}{\degree}, along which it starts to form an antiskyrmion lattice at \SI{-5}{\degree} and enters a mixed non-topological bubble/helical phase. To complete the tilt loop, the sample is finally tilted back to \SI{-7}{\degree}.

Non-topological bubbles, antiskyrmions, and sections of the helical phase were manually identified, marked, and counted to determine their number from the LTEM images. However, counting the total number of antiskyrmions over the whole field of view was not possible due to the unavoidable occurrence of bending contours in the LTEM images that shifted upon tilting and inhibited the visual recognition of some of the texture elements in parts of the image. Consequently, the ratio of antiskymrions (AS) to non-topological bubbles (NT) was used to identify those tilt angles, where antiskyrmions occur. 

\begin{figure}[t]
\includegraphics[width=7.5cm]{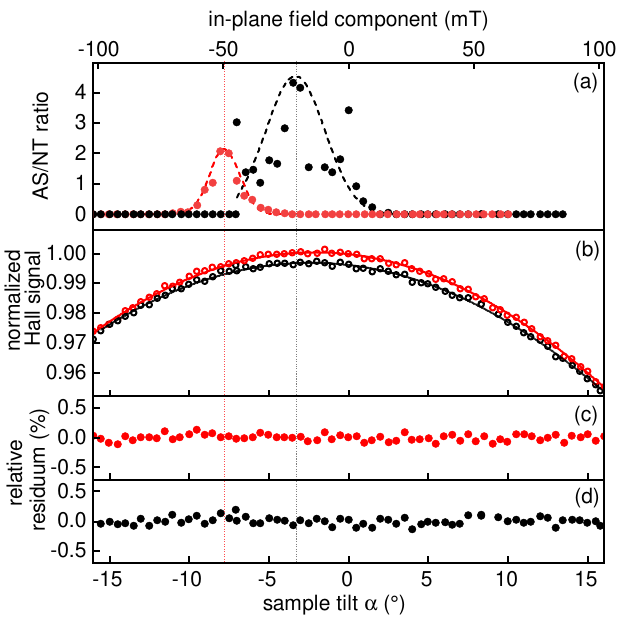}
\caption{(a) Ratio of antiskyrmions (AS) and non-topological bubbles (NT) as function of the tilt angle. The resulting component of the field parallel to the plane of the sample is given on the upper abscissa. The dotted black (forward tilt) and red (backward tilt) lines indicate the angles, where the maximum number of antiskyrmions are observed in LTEM images. (b) Variation of the in-situ measured Hall signal during the same tilt series. The lines indicate cosine fits to the data with a shift of \ang{1.5}. (c,d) A subtraction of these cosine fits from the data points and subsequent normalization yields the relative residuum. No peculiarities are observed upon the occurrence or vanishing of the antiskyrmions.}
\label{fig:Topo-Hall}
\end{figure}

This ratio is plotted as a function of the tilt angle $\alpha$ in Fig.~\ref{fig:Topo-Hall}a, where the color-code represents the different directions of the tilt angle (black: negative to positive angles, red: positive to negative). From this analysis, antiskyrmions are identified to occur in the angular ranges $\SI{-7}{\degree} \leq \alpha \leq \SI{+3}{\degree}$ and $\SI{-5}{\degree} \leq \alpha \leq \SI{-10}{\degree}$ during the up and down tilts, respectively, and are found to be mostly present at approximately \SI{-3.5}{\degree} and \SI{-8}{\degree}. 

Fig.~\ref{fig:Topo-Hall}b shows the simultaneously measured Hall voltage normalized with respect to its maximum value as a function of the sample tilt. The Hall signal exhibits a cosinusoidal drop with increasing (positive and negative) tilt angle, which is expected assuming an originally out-of-plane oriented ferromagnetic background that follows the direction of the applied external field (i.e., $M_{\perp}(H) \propto \cos(\alpha + \Delta\alpha))$. The slight shift of the origin of the cosine function by $\Delta\alpha = \SI{-1.5}{\degree}$ is ascribed to some unavoidable mis-tilt of the lamella imposed upon placing it onto the substrate in the in-situ holder. 

A small offset of the Hall voltage of \SI{25}{nV} is found upon comparing the up and down tilts. An estimation based on the areal density of magnetic features in the LTEM images and the measured Hall voltages shows that this corresponds to exactly that amount of anomalous Hall voltage that is to be added or subtracted, when a single non-topological bubble or antikyrmion is created or annihilated on an otherwise ferromagnetic background. We subtract the respective consine fit for both tilt directions from the data and depict the normalized values, i.e.\ the relative residuum in Fig.~\ref{fig:Topo-Hall}c,d. Most strikingly, we do {\em not} observe any changes or peculiarities in the course of the Hall signal that could be related to the occurrence of antiskyrmions as indicated by the AS/NT ratio in Fig.~\ref{fig:Topo-Hall}a.

Please note that the longitudinal resistance change determined from the voltage drop across the sample between the current contacts was determined to be \SI{0.02}{\%} in the investigated field range and can, thus, be neglected for the interpretation of the transversal Hall voltage, while the relative error due to the limited resolution and accuracy of the measurement equipment and the reproducibility amounts to \SI{5}{nV}, which corresponds to \SI{2.4}{n\ohm cm}.

Before we conclude, let us reiterate the essential details: (i) The measured AHE is proportional to the out-of-plane component of the sample magnetization. (ii) Micromagnetic simulations of antiskyrmions and non-topological bubbles show that both magnetic textures have similar out-of-plane components \cite{Peng:2020}. (iii) Even though completely different magnetic textures such as antiskyrmions or non-topological bubbles are present and transformed into each other, \emph{no impact on the Hall voltage} is observed.


%
In summary, we have looked into the magnetotransport properties in metallic \ch{Mn_{1.4}PtSn}, because this compound is known to host antiskyrmions at room temperature. Utilizing high-quality single crystals, we have prepared micro lamellae to perform concurrent Lorentz-TEM and in-situ Hall investigations on this inverse Heusler compound. Employing a well established magnetic field protocol, we 
 first generated a stable antiskyrmion lattice in the sample, then transformed the antiskyrmions into non-topological bubbles and vice-versa. No additional Hall voltage related to the (dis)appearance of antiskyrmions was found, i.e., the antiskyrmions did not give rise to a topological Hall effect. 

\begin{acknowledgments}
We thank T.\ Helm and R.\ Sch\"afer for valuable discussions on the topic. This research was funded by the Deutsche Forschungsgemeinschaft (DFG, German Research Foundation) within the framework of the priority program SPP 2137 (project ID 403503416). We acknowledge the use of the facilities in the Dresden Center for Nanoanalysis (DCN) at the Technische Universität Dresden. D.P. acknowledges the support bythe DFG through the project 504660779. M.W.\ acknowledges support from the International Max Planck Research School for Chemistry and Physics of Quantum Materials (IMPRS-CPQM). D.K. acknowledges the support by the Lumina fellowship LQ100102201 of the Czech Academy of Sciences and the Czech Science Foundation (grant 22-22000M). S.S.\ acknowledges support from the DFG through the projects 458685885 and 531289024. 
\end{acknowledgments}

\bibliography{references.bib}

\end{document}